# Towards an Infrastructure-less SOA for Mobile Web Service Composition

Chii Chang     Sea Ling
Faculty of Information Technology, Monash University
900 Dandenong Road, Caulfield East,
Victoria 3145, Australia

{chii.chang, chris.ling}@monash.edu

## ABSTRACT
Service composition enables customizable services to be provided to the service consumers. Since the capabilities and the performances of mobile devices (e.g., smart phone, PDA, handheld media player) have improved, a mobile device can be utilized to interact with external mobile service providers towards providing composite Web service to remote clients. Existing approaches on mobile-hosted service composition are usually platform dependent, and rely on centralized infrastructure. Such approaches are not feasible in an open, mobile infrastructure-less environment, in which networked services are implemented using different technologies, devices are capable of dynamically joining or leaving the network, and a centralized management entity is nonexistent. This paper proposes a solution to enable mobile Web service composition in an open, infrastructure-less environment based on loosely coupled SOA techniques.

## Categories and Subject Descriptors
D.2.m **[Miscellaneous]**

## 1. INTRODUCTION
Mobile devices (e.g., smart phones, PDAs, or handheld media players) are not only capable of consuming distributed services, but also able to function as service providers. A mobile host can also perform service composition to support more complex needs. For example, the mobile host can interact with environmental sensor services to share the information of its current environment with a remote service consumer who is located in another location.

There are a number of works [1]-[4] proposed for service composition in mobile service environment. They mainly focused on semantic service selection, or QoS (Quality of Service) based service selection. To distinguish these works from ours, the focus of this paper is on how to enable dynamic service composition in an open, mobile infrastructure-less environment based on loosely coupled SOA (Service Oriented Architecture).

Supporting dynamic service composition in a mobile network can be simply realized if the connected entities in the entire network are implemented using a single technology. Such an assumption is not feasible in an open, mobile infrastructure-less environment, in which not only a central management entity (e.g., service repository) is unavailable, but also each connected mobile peer may be implemented in different technologies. A number of researchers [5]-[8] have proposed solutions based on loosely coupled SOA (Service Oriented Architecture) for such an environment, and adapted XML Web service as the common interface for the network services to enhance the overall interoperability.

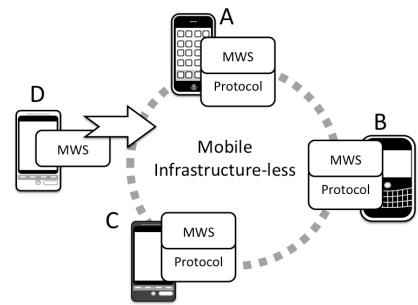

**Figure 1. Mobile infrastructure-less environment using Web services.**

Figure 1 illustrates a mobile infrastructure-less topology, in which there are three mobile peers that are connected to the network using a specific protocol. Each mobile peer applies Web service as the common interface for the interaction. The mobile peer D is a newly joined mobile Web service (MWS) that intends to interact with the other MWSs to perform service composition autonomously with minimal interference. However, the mobile peer D is unable to interact with the other MWSs, because the mobile peer D does not have prior knowledge of this environment (e.g., the IP and capabilities provided by each MWS), and because the environment is infrastructure-less, mobile peer D is unable to get any assistance from a central management entity (e.g., a broker). Hence, autonomous interaction is still not possible even though Web service has been applied as the common interface.

In this paper, we propose a solution that enables dynamic mobile Web service composition in an open, mobile infrastructure-less environment based on loosely coupled SOA techniques, so that problems described in the scenario above can be resolved.

The paper is organized as follow: We review a number of related works in Section 2, and describe our proposed solution in Section 3. Section 4 provides the detail of the prototype implementation. We conclude our work, and describe the future work in Section 5.

## 2. Related Works
### 2.1 Mobile Web Service Provision

A common approach to realize MWS is to build additional Web service related components on top of existing mobile distributed systems. There are various frameworks available to be the foundations of MWS provision, such as Jini [8], SIP [5], OSGi [9], and JXTA [7], [10]. Generally, MWS provision can be classified into centralized, and decentralized.

The centralized MWS provision [5], [8], [9] requires some form of centralized control (e.g., a repository, or a broker) to be made available within the environment, in which mobile peers are capable of publishing/subscribing themselves to the central entity.

The decentralized MWS provision can be further categorized into two approaches: (1) the peer-group based approach [7] requires a higher performance node (super peer) in a group of peers. It is responsible to assist the peers' publish/subscribe. (2) the pure peer-to-peer based approach denotes an open mobile infrastructure-less environment that is purely dynamic, and has no available centralized repository, and neither dose it have super peers to realize the publish/subscribe mechanisms. Our MWS composition solution requires this kind of approach to enable dynamic interaction. However, existing MWS provision solutions [5], [7], [8], [11] do not cover this kind of architecture.

For realizing MWS provision, JXTA appears to be a popular solution in recent years, due to its platform-independent characteristic, which enables various network resources to participate within the network. One limitation mentioned in [12] is that a JXTA peer is unable to communicate with an external peer, which is located in another P2P protocol. To address to this limitation, Srirama et al. [13] have proposed a middleware to enhance the scalability of JXTA framework based MWS environment.

### 2.2 Enterprise Service Integration for Mobile Web Services

Enterprise Service Bus (ESB) is a software infrastructure that can easily connect IT resources by combining and assembling services to achieve a Service Oriented Architecture (SOA). It provides the "Link" between service providers and requesters. A service in ESB is derived from any type of software component implemented in any programming language and on any platform. ESB provides a runtime environment enabling these software components to be registered as manageable ESB service providers – Bus Service Providers (BSP), which are usually utilized as XML Web services, with common service description format (usually in XML format such as WSDL). A common ESB usually encompasses three basic elements [14]:

- Registry – An ESB registry enables service discovery and management of meta-information, which describe the interface, policies and behaviors of a component, enabling the component to be registered as BSP, which can be discovered by the semantic service routing and matchmaking mechanisms.

- Links – It denotes the interaction between BSPs and BSRs (Bus Service Requesters) with policy attachments defining both entities' requirements and criteria. A Link usually requires the support mechanism provided by Mediation.

- Mediation – The basic functions provided by Mediation include: managing request and response messages such as redefining the request message for semantic match-making and trans-coding message format in order to route the request message to the BSP; observing the interaction messages and providing certain level management. An ESB usually adapts multiple mediations in order to support the interoperability happening in different protocols and handling different criteria. The design of Mediation is differentiated depending on the complexity of the entire system.

Enterprise Service Bus (ESB) has been applied to MWS in the work of Mobile Web Service Mediation Framework (MWSMF) [13], [15]. MWSMF supports the basic ESB features for routing normalized messages between components and end-points. It enables remote clients interacting with the P2P network peers via standard protocol (e.g., HTTP). However, it does not support dynamic service composition and neither does it support dynamic service interacting in a mobile infrastructure-less environment (the "pure p2p" mentioned in their work [7]). Furthermore, MWSMF was proposed for the peer-group based approach, but the aim of our research is for pure peer-to-peer.

## 3. Mobile Web Service Composition Mediator

In this section, we describe our proposed solution to enable dynamic service composition in the mobile infrastructure-less environments. Distinguishing our work from others that focus on the service selection and service reasoning based on QoS or semantic domain [1]-[4], our solution mainly focuses on modeling the generic architecture towards realizing the loosely coupled SOA based on the ESB architecture, which we called Mobile Web Service Composition Mediator (MWSCM) - a mobile-hosted composite Web service using ESB based middleware.

### 3.1 Architecture Overview

Figure 2 illustrates the proposed architecture for MWSCM. It follows the basic ESB architecture that lies in the middle to realize the "link" between the service provider and the service requester. The mediator combines with additional components to support the on-the-fly dynamic service interoperability in the mobile infrastructure-less environments. We describe the components of the architecture below.

MWS Providers, which include the SOAP-MWS Provider, the REST-MWS Provider, and the Socket Service Provider, represent the network resources that are being hosted on the remote devices, and have been advertised in the wireless network. These MWS providers can be implemented in any form, either a simple socket service provider that communicate over the TCP/IP or UDP/IP protocol, or a SOAP or RESTful [16] service provider that communicates over the HTTP protocol. These service providers apply Web service as their common interface, in which a common service description format – WSDL 2.0[1] is used to describe the functionalities of each MWS. Furthermore, each service description document can contain additional information to describe what type of service it is, and what types of operations it provides. The service description may involve semantic service reasoning, which is not in the scope of this paper. We recommend [17] for further reading in such research domain.

---

[1] http://www.w3.org/TR/wsdl20/

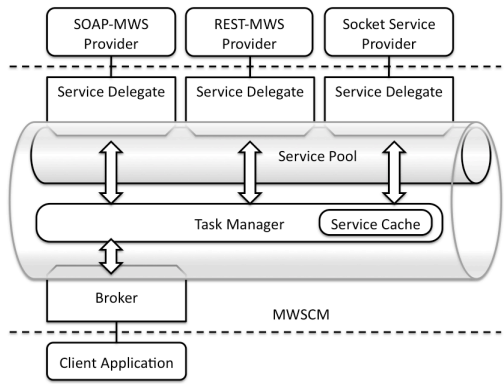

**Figure 2. MWSCM Architecture.**

A MWS is advertised, and is discoverable in an open mobile infrastructure-less network environment. Such an environment in related works was realized in a platform-dependent manner [8], or centralized model [5]. Our solution requires loosely coupled SOA, in which the environment must be implemented based on a platform and language independent open protocol. There are a number of technologies capable of supporting such a need: e.g., JXTA[2], UPnP[3], Bonjour[4], or Avahi[5].

Service Pool (SP) enables the "link" feature of ESB by creating Service Delegates (SD) as links to service providers. SP has two main functions:

- Net Service Browsing – SP browses the entire network continuously when MWSCM is launched, until it is commanded to stop. The name or the URI of the discovered service is stored on the runtime service list instance. The service list is updated when a new service joins the network, or when a service leaves the network.

- Service Registration – SP also allows service providers to register themselves to the mediator. This mechanism is used in a user-driven environment, in which the mediator user (a newly joined mobile peer to the network) allows other MWSs to automatically interact with the user without user to perform the discovery process.

SP allows other components to search the service list by passing the Service Type parameter, or by passing the Operation Type. It runs a loop to obtain the service description document of each service provider to define which service provider matches the need. A matched service provider will have its SD created. The resulting SD list will then be passed back to the requester.

A Service Delegate (SD) is a runtime instance that is dynamically created by the SP based on the service description document (e.g., WSDL) of the service provider. A SD acts as the proxy of its linked service provider endpoint. It is created when a service provider is identified as the feasible service for a task (which is identified by the Task Manager (TM) when the TM searches the services in Service Pool). The SD enables the mediator's interaction with a linked service provider endpoint just like the way it interacts with an internal component.

Task Manager (TM) enables intelligent routing based on user-defined criteria. Each normalized request sent from the Broker is to be analyzed by matching the task to the task organization document. A task organization document is a well-formatted document that describes what operations are required to accomplish the request. For example, the requester would like to know the absolute location of the mediator user while the mediator user is inside a building. Such a request requires the mediator to interact with the global positioning service and the indoor location positioning service. Based on the requirement, TM requests the service list based on the operation type (e.g., positioning) from SP.

In the lifetime of a request handling process, in which a number of tasks may be involved, an external service provider may be invoked in multiple tasks. Hence, TM uses a runtime instance — Service Cache (SC) to store the SDs corresponding to the service providers that are associated to a task. SC reduces the need of discovering services from SP, in which the discovery process requires retrieving service description from the service provider, and parsing the service description.

The broker is mainly responsible for normalizing the request data, and formatting the response data for the client application based on the type of the application when the result is returned from the TM. The request sent by the client application, and the expected response data for the client application can be in various forms such as a simple HTTP string, an XML-formatted data, a JSON formatted data, or an encoded data, etc. The exact functionality of the Broker is distinguished by the complexity of the system.

The Client Applications of mediator is the request receiver. It can be an internal component that provides a gateway to enable the internal graphical-user-interface-based application to use the mediator directly, or it can also be a mobile Web service that enables remote clients (e.g., a simple HTTP web application in another mobile user's device) to interact with the environmental service via the mediator user's device.

### 3.2 Interoperation Details

In this section, we explain how the mediator handles each request sent from the client application.

Figure 3 illustrates a simple request-handling scenario in which a request has been sent from the user application to the mediator. We explain the details of the process below:

1. A request has been submitted to the MWSCM Client Application (CA) via the protocol it supports.

2. The CA passes the request data to the Broker to normalize the request format.

3. The Broker normalizes the request data to the common format used by the mediator.

4. Once the request data has been normalized, it is passed to the Task Manager (TM).

5. TM analyses the request by matching it to the predefined task organization document. Based on the number of operations defined in the task organization document, the following steps 6 to 15 may be repeated for each operation, until each operation is completed.

---

[2] https://jxta.dev.java.net/

[3] http://www.upnp.org

[4] http://developer.apple.com/opensource/internet/bonjour.html

[5] http://avahi.org/

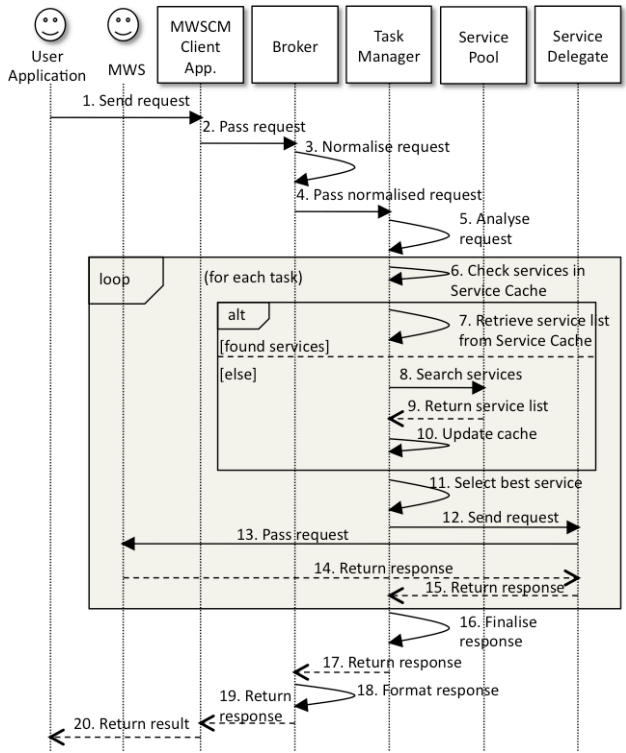

**Figure 3. The sequence diagram of the request handling.**

6. TM checks Service Cache (SC) for a list of services that are capable of performing the give operation type. If the corresponding services are found, step 7 is performed. Otherwise, step 8 will be performed. Note that in current step, and the related steps 8, the request for retrieving the service list is based on the operation type. The operation type is a type defined in a tree-based model. Such a tree-based model has been documented in an XML-formatted file, and has been used as a common description for the service operations of each service provider. A more specific example is the ontology document, which models the service and the operation of service using semantic service description technique.
7. TM retrieves the service list from the service list cache.
8. TM requests SP for a list of services that are capable of performing the given operation type.
9. SP returns a list of SD to TM.
10. TM updates SC.
11. After the TM receives the service list (which is a list of SDs), it compares the status of each service to identify the best service provider.
12. TM sends the request to the SD.
13. SD passes the request to its linked service provider.
14. The linked service provider returns the response to SD.
15. SD passes the response back to the TM.
16. Once the response of each operation is returned back to the TM, it finalizes the result data.
17. TM returns the finalized result data back to the Broker.
18. Broker parses the result data to the format which the CA expects to receive.
19. Broker returns the formatted result data back to the CA.
20. CA sends the final response to the user application.

## 4. Prototype Implementation

This section provides the details of our proof-of-concept prototype implementation. We describe how the MWS is realized, how the MWSCM is implemented, and what technologies have been applied to realize the mobile infrastructures environment for dynamic service interaction.

### 4.1 Implementation Setup

We have implemented the prototype by using a number of mobile devices; including an Apple iPhone 3GS, an Apple iPod Touch, and an Apple MacBook, which runs the iPhone emulator. The implementation was developed in the Xcode IDE [6] using Objective-C programming language. The implementations run on iPhone OS 3.0. The MWS was built on top of CocoaHTTPServer – iPhone version [7]. CocoaHTTPServer supports the Bonjour-enabled socket service to be hosted on the iPhone, and communicable via TCP/IP, UDP/IP, or HTTP. On top of CocoaHTTPServer, we added additional components to enable the standard XML Web service mechanisms. In our prototype, the MWSs were implemented as RESTful Web services [16]. Since iPhone OS does not fully support the functionality of XML document processes, we have used the GDataXML components from Google's GData[8] package to support the XML document processes.

Figure 4 illustrates the architecture of the prototype implementation. There are three main entities in this prototype: the client application, the MWSCM host, and the MWS provider. Note that a mobile device in such an environment can act as both service provider and requester. We briefly describe the implementation of each entity:

*MWS provider* is a Web service built on top of CocoaHTTPServer and hosted on a mobile device. Its service is discoverable within the Zeroconf[9] network. Note that we used Bonjour as the Zeroconf implementation.

*MWS Requester* is a simple Web service client application with a component for browsing services in the Zeroconf network. It is an HTML-based application for user to submit a request to the mediator host.

The *mobile host* shows in the middle of Figure 4 is the mediator host. The MWSCM and a Client Application are hosted on this device. As the figure shows, the Client Application is a Web service that handles request and response for the remote client application. This Web service has been implemented as a RESTful Web service. Hence, it can easily accommodate interactions with by various client applications. The MWSCM follows the design based on ESB architecture that has been described in Section 3. The Net Service Browser component has been implemented as a DNS-SD (DNS – Service Discovery) component that is capable of discovering service advertisements in the Zeroconf network.

---

[6] http://developer.apple.com/tools/xcode/
[7] http://code.google.com/p/cocoahttpserver/
[8] http://code.google.com/p/gdata-objectivec-client/
[9] http://www.zeroconf.org/

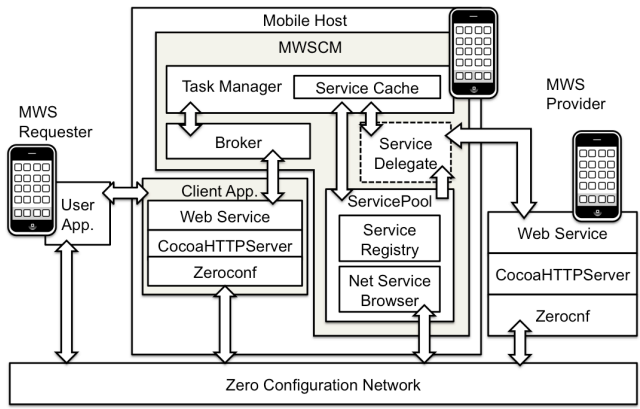

**Figure 4. The architecture of prototype.**

## 4.2 Test Case

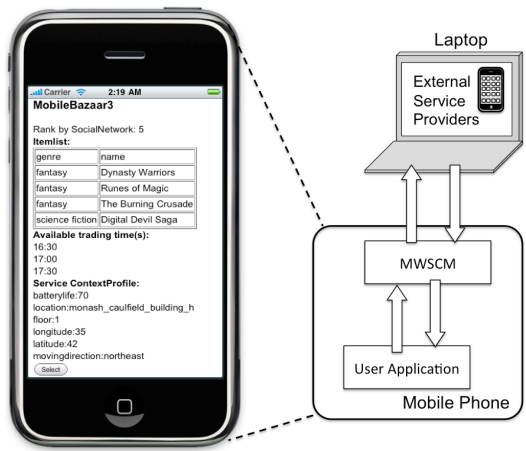

**Figure 5. The recommended MBV.**

Our test case is based on a scenario in which a mobile user connects his/her device to the virtual social network. The user is trying to find a virtual Mobile Bazaar Vender (MBV) who is selling DVDs. An MBV is a mobile Web service that provides the information about what items are available to be purchased from the device owner. Since there are numerous MBVs on the network, the user decides to request the MBV broker (broker) for a recommended MBV. The broker is a MWS embedded with MWSCM. It receives the request from other mobile users, and retrieves contents from the other MBVs based on the requester's need (Note that the broker can also be hosted on the user's device). The user then sends the request to the broker together with his/her preference profile, which describes what his/her favorite DVD genres are, and the preferred trading times for him. The broker interacts with the other MBVs based on the algorithm described in Section 3.2. When the broker finalizes the result, it sends the information of the recommended MBV back to the user's device.

The left hand side of Figure 5 shows the screenshot of the user's device with the final result displayed.

## 4.3 Performance Evaluation

We have evaluated the performance of the service composition process in two experiments, and both have been conducted within a Wi-Fi network environment with data rate at 54 Mbps. The MWSCM was hosted on an iPhone 3GS, and a number of MWS providers were deployed on the iPhone simulator of a Macbook (2.4 GHz CPU performance, and 4GB RAM). On the right hand side of Figure 5 illustrates the relationship of involved devices in our experiments.

The first experiment aims to evaluate the time spent for a request, which is affected by the number of matched MWS providers in the environment. In this setting, the MWSCM receives a request that needs to be completed with four tasks. Each task is performed using the algorithm described in Section 3.2. We measure the time spent at four different settings in terms of the percentage of tasks using SC. Figure 6 illustrates the result of the first experiment.

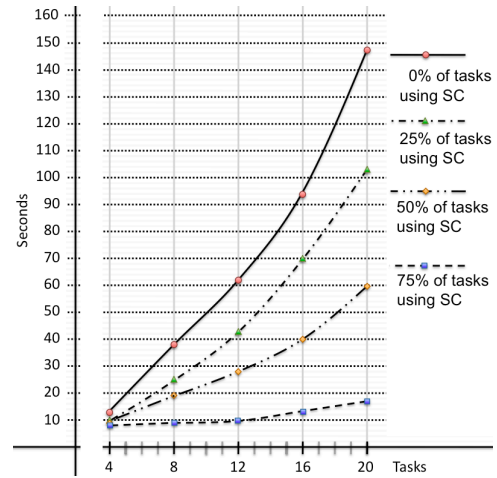

**Figure 6. Time affected by number of tasks.**

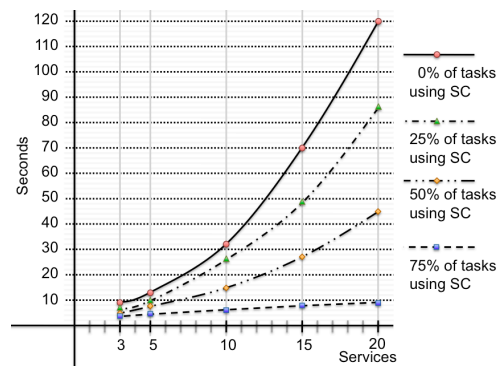

**Figure 7. Time affected by number of services.**

The second experiment aims to evaluate the overall performance affected by the number of required tasks (i.e. invoking service) to complete a request. We have deployed five MWS providers on the Macbook, and have requested the MWSCM to perform four tasks repeatedly to simulate the time spent affected by the number of tasks. We also measure the time spent at four different settings in terms of the percentage of tasks using SC. Figure 7 illustrates the result of the experiment.

The experiments demonstrate that the overall performance can be improved when Service Cache is used.

## 5. Conclusion and Future Works

In this paper, we have described our proposed solution for a dynamic mobile Web service composition in an open, mobile infrastructure-less environment based on loosely coupled SOA architecture. The solution applied ESB architecture that enables dynamically discovered MWSs to be utilized as pluggable components. We have implemented a prototype as proof-of-concept, and have tested the prototype by applying a test case based on a composite MWS scenario. Furthermore, the performance of MWSCM in terms of cache usage has been evaluated.

We intend to extend our solution, and include the following features in our future work:

- Context/Resource-aware service caching – we intend to adapt context and resource awareness to enhance the mechanism of service caching.
- Adaptive interoperability – the MWSCM will observe each interacting peer to ensure the selected service provider can provide a stable communication.
- Failure detection and recovery – MWS composition can fail because of numerous mobility-related issues, e.g., the MWS provider is suddenly switching its connection from WiFi to 3G. MWSCM should provide mechanisms to support failure detection and recovery.